\documentclass[superscriptaddress,aps,prx,reprint,twocolumn,longbibliography,9pt]{revtex4-2}
\usepackage[utf8]{inputenc}
\usepackage{amsfonts}
\usepackage{physics}
\usepackage{graphicx}
\usepackage{overpic}
\usepackage{color}
\usepackage[dvipsnames]{xcolor}
\usepackage{comment}

\definecolor{myblue}{HTML}{0070de}

\usepackage[pdftex, colorlinks=true, pdfstartview=FitV, linkcolor= black, citecolor= myblue, urlcolor= myblue, hyperindex=true,hyperfigures=true]{hyperref}

\usepackage[T1]{fontenc} 
\usepackage{enumitem}
\usepackage{booktabs}
\usepackage{float}

\usepackage{soul}

\usepackage{orcidlink}
\newcommand{\orcidalice}{\orcidlink{0000-0002-0563-5174}}
\newcommand{\orciddaniel}{\orcidlink{0000-0001-7658-3546}}
\newcommand{\orcidsimone}{\orcidlink{0000-0002-8882-2169}}
\newcommand{\orcidwerner}{\orcidlink{0000-0001-7199-0978}}

\newcommand{\uulm}{Institute for Complex Quantum Systems, Ulm University, Albert-Einstein-Allee 11, 89069 Ulm, Germany}
\newcommand{\unipd}{Dipartimento di Fisica e Astronomia "G. Galilei" \& Padua Quantum Technologies Research Center, Universit{\`a} degli Studi di Padova, Italy I-35131, Padova, Italy}

\newcommand{\pdinfn}{INFN, Sezione di Padova, via Marzolo 8, I-35131, Padova, Italy}

\usepackage{lineno}

\begin{document}

\title{Optimal control transport of neutral atoms in optical tweezers at finite temperature}

\author{Alice Pagano\orcidalice}
\affiliation{\uulm}
\affiliation{\unipd}
\affiliation{\pdinfn}

\author{Daniel Jaschke\orciddaniel}
\affiliation{\uulm}
\affiliation{\unipd}
\affiliation{\pdinfn}

\author{Werner Weiss\orcidwerner}
\affiliation{\uulm}

\author{Simone Montangero\orcidsimone}
\affiliation{\uulm}
\affiliation{\unipd}
\affiliation{\pdinfn}


\begin{abstract}
  The transport of neutral atoms in Rydberg
  quantum computers is a crucial step of the initial arrangement
  of the grid as well as to the dynamic connectivity, recently successfully
  demonstrated.
  We study the application of optimal control and the quantum speed limit for the transport of neutral
  atoms in optical tweezers at finite temperatures and analyze how laser noise affects transport fidelity. Open-loop optimal control significantly enhances transport fidelity, achieving an improvement up to 89\% for the lowest analyzed temperature of $1\,\mu$K for a distance of three micrometers.
  Furthermore, we simulate how the transport fidelity behaves in release-and-capture measurements,
  which are realizable in the experiment to estimate transport efficiency and implement closed-loop optimal control.
\end{abstract}

\maketitle


Neutral atom quantum processing units based on optical tweezers~\cite{Saffman2018, Henriet2020, Keijzer2023, Wilson2022, Graham_2022, Wintersperger2023, Norcia2024} open the
possibility to modify the connectivity of qubits at runtime of the algorithm. In the pioneering experiments in Ref.~\cite{Bluvstein2022, Bluvstein2023}, coherent transport of atoms through the movement of optical tweezers has been demonstrated.
Ensuring high-fidelity atom movement is crucial not only for enabling non-local connectivity but also for the necessary reshuffling of atoms in the initial array arrangement, after the stochastic loading of atoms within the optical tweezers~\cite{Tian2023, Barredo2016, Endres2016, Mello2019, Norcia2018, Kaufman2012}.
A fast high-fidelity transport can be achieved via quantum optimal control (QOC)~\cite{Caneva2009, Zhang2018, Koch2022, Furst2014}.
Optimal control has emerged as an essential tool for optimizing quantum gates as well~\cite{Levine2019, Jandura2022, Pagano2022, Bluvstein2023}.
QOC allows for either open-loop optimization using numerical simulations or closed-loop optimization directly integrated with the experimental setup.
A variety of optimal control methods is available via open-source package~\cite{Rossignolo2023} as the dressed chopped random basis (dCRAB) algorithm~\cite{Caneva2011, Rach2015, Muller2022}.

\begin{figure}[t]
    \centering
    \includegraphics[width=\linewidth]{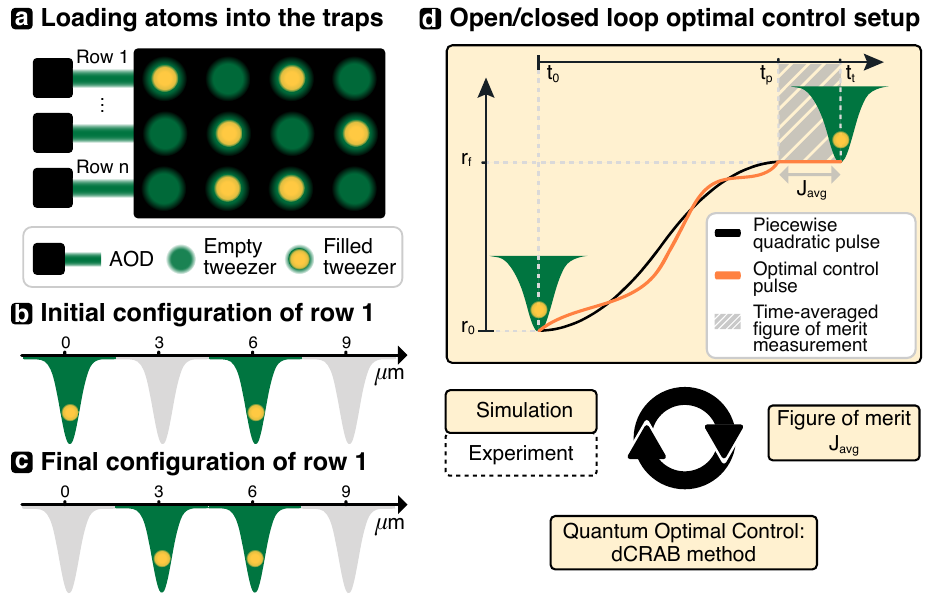}
    \caption{\textbf{Setup for atoms transport.}
        (a)
        An array of neutral atoms after the initial stochastic loading. Empty traps are switched off and atom transport along one row of the array is possible via optical tweezer generated by acousto-optical deflectors (AODs).
        (b-c)
        We aim to move the atom
        at position $r_0 = 0 \mu \mathrm{m}$ to the position $r_{\mathrm{f}} = 3 \mu \mathrm{m}$, a vacant spot in the example.
        (d)
        We perform open-loop optimal control via the time-averaged figure of merit $J_{\mathrm{avg}}$; closed-loop optimal control is anticipated by taking into account additional experimental constraints in our simulation, especially on the pulse shape $r(t)$ that controls the position of the trap.
    \label{fig:setup}}
\end{figure}

We design an open-loop optimal transport of neutral atoms via optical tweezer
at finite temperature to closely replicate experimental conditions: indeed, in typical experiments, atoms are trapped in
Gaussian-shape potential formed by optical tweezers and each atom can be
represented as a thermal state at a temperature approximately around $10\mu \mathrm{K}$~\cite{Hoelzl2023}.
After the stochastic loading of the atoms, empty traps are switched off, as depicted in Fig.~\ref{fig:setup}(a). We have control to move tweezers within a given row through acousto-optical deflectors (AODs). We reshuffle the atoms to occupy the vacant positions, thereby achieving the desired layout. 
The movement must be executed at sufficient fidelity and speed to prevent loss and avoid any temperature alterations~\cite{Torrontegui2011}.
We take into consideration a variety
of experimental constraints for optical tweezers implemented via acousto-optical
deflectors, e.g., a finite frequency bandwidth, and piecewise quadratic pulse
shapes, both motivated by the experiment in Ref.~\cite{Hoelzl2023, Unnikrishnan2024, Meinert2021}. We also introduce noise into the modeling of the tweezer trap, i.e. for its depth, waist, and position.
We find that, with our setup, the quantum speed limit for moving a Strontium-88 atom over a distance of three micrometers is around $10\,\mu$s. The quantum speed limit relates to the minimum time required to accomplish a physical process. For transport, we extrapolate this as the time when it is no longer possible to improve the figure of merit, in agreement with~\cite{Lam2021}.
Furthermore, at the lowest analyzed temperature, a notable enhancement in the figure of merit is observed, reaching a factor of 89\%.
We examine the impact of laser noise on the trap on the figure of merit.
The release-and-capture protocol is the standard approach
to verify the quality of the atom cooling~\cite{Hoelzl2023}, which also 
applies to the evaluation of the heating during transport.
We predict which level of precision of the figure of merit can be observed
experimentally by running the release-and-capture simulations after the optimal pulses.
The proposed analysis applies only to the initialisation of the array. Extending it to mid-circuit movement of atoms would require encoding the internal degrees of freedom of the atoms and decoherence effects, which is beyond the scope of this work.

We first describe the open-loop optimal control setup in Sec.~\ref{sec:setup}.
Then, we turn to the analysis of the quantum speed limit in Sec.~\ref{sec:qsl}, followed by simulations with noise and an explanation of the results in release-and-capture measurements based on these simulations.
We conclude the results in Sec.~\ref{sec:conclusion}.

\section{Optimal control setup                                     \label{sec:setup}}

We focus on the general problem of moving one atom from position $r_0$ to $r_\mathrm{f}$, e.g., from the configuration shown in Fig.~\ref{fig:setup}(b) to that in Fig.~\ref{fig:setup}(c).
Each atom is in a thermal state
\begin{align}
   \rho(T) &= \sum_{i=1}^{N_s} p_{i} \ket{\psi_{i}} \bra{\psi_{i}} \, ,
   \label{eq:thermal-state}
\end{align}
where $p_{i} = p_{i} (T) = \mathrm{e}^{-E_i/k_b T}$ are the Boltzmann weights at temperature $T$ with $E_i$ the energy of the $i$-state. We introduce a cutoff in the number of states $N_s$ to represent the finite temperature density matrix.
The convergence of the results with the cutoff $N_s$ is discussed in appendix ~\ref{appendix:code}. We model the optical tweezer traps as Gaussian potentials of the following form:
\begin{align}
    V(t) = U_0 \mathrm{e}^{ \frac{- 2 (x - r(t))^2}{w_0^2}  } \, ,
\label{eq:gaussian-trap}
\end{align}
where $U_0$ is the trap depth, $w_0$ the beam waist and $r$ the time-dependent center of the trap~\cite{Endres2016,Graham_2022,Unnikrishnan2024}.
In the experiment, the pulse $r(t)$ tuned to move the atom is generated via arbitrary waveform generators (AWG), which imposes constraints on the available pulse shapes, such as the piecewise quadratic shape~\cite{SpectrumInstrumentation}.
Additionally, we consider laser noise effects and finite bandwidth during the optimization process.
We describe the numerical simulation for computing the time-evolution of the thermal state under time-dependent pulses in appendix~\ref{appendix:code} and integrate quantum optimal control using the open-source suite QuOCS~\cite{Rossignolo2023}.
We employ the dCRAB algorithm~\cite{Muller2022, Rach2015}: the key idea is to construct a pulse $r(t)$ as a sum of randomly assembled basis elements $r_i(t)$ to minimize a given figure of merit $J$. 
The optimization involves different superiterations, each simultaneously optimizing $N_c$ coefficients of the basis.
The full pulse at the superiteration $j$ can be written as
\begin{align}
    r^j(t) = c_0^j r^{j-1}(t)+\sum_{i=1}^{N_c} c_i^j r_i^j(t) \, .
\end{align}
where $r^{j-1}$ is the converged solution of the previous superiteration and $\{c_i\}_{i=0}^{N_c}$ is the set of coefficients to optimize.
A common selection for basis decomposition is the use of trigonometric functions, specifically the Fourier basis; however, the choice of bases can vary depending on the nature of the control problem, with certain bases proving more effective than others~\cite{Pagano2024}.
Specifically, we employed the sinc basis for optimizing the pulse $r(t)$.
The figure of merit is the infidelity between the density matrices at the end of
the pulse $\rho_\mathrm{f}(r)$ and the density matrix of the initial finite-temperature state shifted
to the final position $\rho_0 (r_\mathrm{f})$, i.e.,
\begin{align}
    J (r) = 1 - \Tr(\sqrt{ \sqrt{\rho_\mathrm{f}(r)} \rho_0 (r_\mathrm{f}) \sqrt{\rho_\mathrm{f}(r)} })^2 \, .
    \label{eq:figure-of-merit}
\end{align}
This metric is a reliable distance measure between any two states~\cite{Jozsa1994, Basilewitsch2020}.
A worse figure of merit indicates that at the end of the transport, the atom has some residual kinetic energy, which directly relates to an increase in the atom's temperature. This statement holds as the atoms starts in a thermal state and its temperature can only increase during a unitary evolution.

In the following analysis, we target Strontium-88 atoms which have a mass of $m=1.46 \times 10^{-25} \, \text{kg}$ and we fix a distance between the traps in the array of $d = 3\mu\text{m}$. The analysis for different distances is discussed in Appendix ~\ref{appendix:transport_distances}.
For the Gaussian-shaped trap, we assume a trap depth of $U_0 = -1 \text{mK}$ with a beam waist $w_0 = 0.5 \mu\text{m}$.
The coordinate in position-space is set via the frequency of the AOD, where the conversion is $3 \mu\text{m} = 1$MHz.
The noise for the position, waist, and depth of the Gaussian potential is modeled as follows:
the amplitude of the laser controlling the trap depth exhibits a noise $\delta S$ with a
spectrum following a power-law distribution given by $S(f) = A_{L} / f^{1/2}$, i.e., the relative intensity noise (RIN). We know from measurement~\cite{InternalDiscussion} that
the amplitude is about $A_{L} = 10^{-11}$.
The noise of the laser
itself is not significant for the transport, but the beam is typically modified by fiber couplings, the optical system, and especially the AOD; the AOD
with its raise time of $6 \mu$s modifies the beam.
This is considered as additional noise $\delta U_0$ in the trap depth, introduced by random sinusoidal fluctuations with an amplitude of 1\% and random frequencies below $100\,$kHz, as well as random sinusoidal fluctuations with an amplitude of 5\% and random frequencies exceeding $100\,$kHz.
For the trap waist, we assume sinusoidal noise $\delta \omega_0$ with a frequency of $1/6\,\mu$s motivated by the raise time of the AOD and an amplitude of 1\%. Regarding the trap position, we assume sinusoidal noise $\delta r$ with an amplitude of $0.01\,\mu$m and random frequencies ranging between $50-150\,$MHz.

\section{Time-optimal pulse for atom transport                                       \label{sec:qsl}}

\begin{figure}[t]
    \centering
    \includegraphics[width=\linewidth]{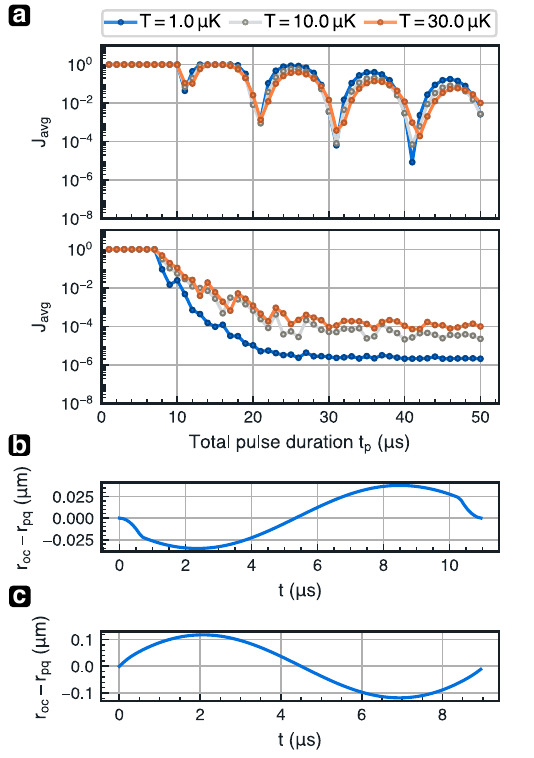}
    \caption{\textbf{Figure of merit as a function of transport time for different temperatures.}
    (a) On top, for a fixed pulse duration $t_{\mathrm{p}}$, the center of the Gaussian-shaped trap is moved via a piecewise quadratic pulse $r_{{\mathrm{pq}}}$ for the distance of $3 \mu$m. We observe periodic minima with a frequency of $2\tau = 10\,\mu$s. When these minima occur, the pulse length $t_{\mathrm{p}}$ is precisely tuned to enable the wavefunction to complete its oscillation period.
    At the bottom, the pulse $r_{{\mathrm{oc}}}$ is obtained with optimal control with $r_{{\mathrm{pq}}}$ as initial guess.
    The optimal control removes the dependency on the wavefunction oscillations and reaches a figure of merit $J_{\mathrm{avg}}$ which monotonically decreases with the pulse time $t_{p}$.
    We observe the quantum speed limit around $2\tau$. The figure of merit depends on the temperature $T$, i.e., higher temperatures result in a
    worse $J_{\mathrm{avg}}$ for the optimal pulse.
    (b) Difference between the optimized pulse $r_{\text{oc}}(t)$ and the initial guess $r_{\text{pq}}(t)$ for $t_\text{p}=11\,\mu$s at $T=1\,\mu$K. (c) The difference is bigger for $t_\text{p}=9\,\mu$s at the same temperature.
    }
    \label{fig:time-vs-fom-finite-T}
\end{figure}

First, we assess the performance of a pulse $r_{\mathrm{pq}}(t)$ matching the AWG's constraint, specifically a piecewise quadratic pulse of the form:
\begin{equation}
r_{\mathrm{pq}}(t) =
\left\{
\begin{aligned}
   & r_0 + 2 \frac{r_\mathrm{f} - r_0}{t_\mathrm{p}^2} t^2 & \quad & t \leq t_\mathrm{p}/2 \\
   & r_\mathrm{f} - 2 \frac{r_\mathrm{f} - r_0}{t_\mathrm{p}^2} (t-t_\mathrm{p})^2 & \quad & t > t_\mathrm{p}/2
\end{aligned} \right.
\end{equation}
where $t_\mathrm{p}$ is the duration of the pulse. Since the figure of merit defined in Eq.~\eqref{eq:figure-of-merit} oscillates at the end of the transport, we evolve the system for an additional fixed $t_\mathrm{c} = 10\,\mu \text{s}$ to compute the time-average $J_{\mathrm{avg}}$.
Thus, the evolution time is $t_\mathrm{t} = t_\mathrm{p} + t_\mathrm{c}$. The optimization workflow is summarized in Fig.~\ref{fig:setup}(d), with an example of a piecewise quadratic pulse represented by the black line.
We examine the time-average of the figure of merit, $J_{\mathrm{avg}}$, as a function of the pulse duration $t_\mathrm{p}$ for the
piecewise quadratic pulse, considering various temperatures for the thermal state in Eq.~\eqref{eq:thermal-state}, specifically for $T=1\,\mu\text{K}, 10\,\mu\text{K}, 30\,\mu$K.
The results are shown in Fig.~\ref{fig:time-vs-fom-finite-T}(a).
The eigenstates of the Gaussian potential lead to typical frequencies and timescales
in the system, which is shown as periodic minima in the figure of merit; this result
agrees with the observation for transport in optical lattices, see
Ref.~\cite{Lam2021}. For the strontium-88 atom and for the selected
trap parameters, the periodicity is given by $2 \tau \approx 10 \,\mu$s with
\begin{equation}
    \tau = \frac{2\pi}{(E_1 - E_0)},
    \label{eq:tau}
\end{equation}
where $E_1$ and $E_0$ are the eigenenergies of the first excited and ground state of
the Gaussian trap, respectively.
Using the piecewise quadratic pulse, we achieve a figure of merit $J_{\mathrm{avg}} < 10^{-2}$ for $t_\mathrm{p} \ge 20\,\mu$s, for the
considered temperatures as shown at the top of Fig.~\ref{fig:time-vs-fom-finite-T}(a). We define the first transport time that satisfies this condition on $J_{\mathrm{avg}}$ as $t_{\text{min}}$. For $T=1\,\mu$K, we have $t_\text{min}^{\text{pq}}=20\,\mu$s.

\begin{figure}[t]
    \centering
    \includegraphics[width=\linewidth]{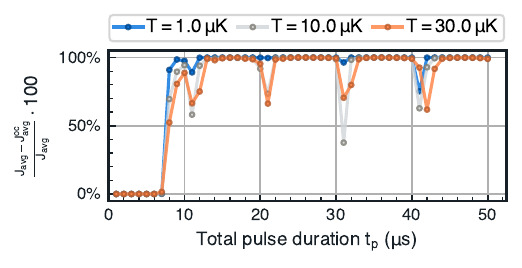}
    \caption{\textbf{Relative improvement of optimal control solution compared to piecewise quadratic pulse.}
    We observe a substantial improvement with optimal control. This enhancement is particularly significant for pulses $r_{\mathrm{oc}}$ with a duration $t_{\mathrm{p}}$ aligning with minima observed in the piecewise quadratic solution $r_{\mathrm{pq}}$.
    Within the minima, we are interested in the region around the quantum speed limit.
    For example, we observe a significant advantage at 
    $t_{\mathrm{p}}=11\,\mu$s, where the figure of merit improves
    by $89\%, 58\%, 66\%$ for $T=1\,\mu\text{K}, 10\,\mu\text{K}, 30\,\mu$K, respectively.
    }
    \label{fig:fom-oc-ratio}
\end{figure}

At the bottom of Fig.~\ref{fig:time-vs-fom-finite-T}(a) we show the results of the optimization of the pulse shape via QOC: we analyze $J_{\mathrm{avg}}$ as a function of the optimal pulse with duration $t_{\mathrm{p}}$. The
initial guess for the optimization is the piecewise quadratic pulse $r_{\mathrm{pq}}$.
This data shows the typical feature of the quantum speed limit in an optimal control problem~\cite{Beugnon2007,Lam2021}: we extract the quantum speed limit from the bottom plot of Fig.~\ref{fig:time-vs-fom-finite-T}(a) as $2 \tau$ where the figure of merit does not improve anymore. This estimate is in good agreement with the timescale of the gap in Eq.~\eqref{eq:tau}.
From now on, we interchangeably refer to the quantum speed limit or $t_\text{min}$, the minimum time where the condition $J_{\mathrm{avg}} < 10^{-2}$ holds. With optimal control and for $T=1\,\mu$K, we have $t_\text{min}^{\text{oc}}=11\,\mu$s. We note that this estimate is almost half the one with the piecewise pulse.
In Fig.~\ref{fig:time-vs-fom-finite-T}(b), we show the difference of an optimized pulse $r_{\text{oc}}$ with $t_\text{p}=11\,\mu$s with its initial guess.
In this case, the optimal control solution adds a small correction to the piecewise quadratic shape to match the wavefunction's periodicity. 
Another example is shown in Fig.~\ref{fig:time-vs-fom-finite-T}(c).
In general, the optimal control optimization eliminates the figure of merit oscillations that were present in the piecewise quadratic case and results in $J_{\mathrm{avg}}=10^{-2}$ at around $t_{\mathrm{p}}=10-12\,\mu$s depending on the temperature $T$.
We observe that the temperature significantly influences the figure of merit for atom transport. Higher temperatures of the initial state lead to a worse figure of merit for the optimal pulse, with no room for further optimization.
Moroever, we emphasize that the quantum speed limit depends on the fixed trapping parameters and we verify that it increases for a shallow trap. For instance, if we fix as trap depth $U_0=-0.5\,$mK, the quantum speed limit is $t_{\text{min}}^{\text{oc}}=14\,\mu$s.
Surprisingly, for deeper traps the minimum time does not decrease but it converges to $t_{\text{min}}^{\text{oc}}=11\,\mu$s.
The relative improvement of the optimal control solution for the piecewise quadratic pulse is illustrated in Fig.~\ref{fig:fom-oc-ratio} as a function of the pulse duration.
The improvement holds particular significance at pulse duration $t_{\mathrm{p}}$ where minima were evident for $r_{\mathrm{pq}}$ and as close as possible to the quantum speed limit.
Specifically at $t_{\text{min}}^\text{oc} = 11\,\mu$s, the figure of merit can be improved by a factor of $89\%, 58\%, 66\%$ for $T=1\,\mu\text{K}, 10\,\mu\text{K}, 30\,\mu$K, respectively.
Furthermore, we verify that a comparable improvement is observed when we impose the experimental constraint on the pulse shape fitting it with piecewise quadratic segments, each spanning $10\,\mu$s.

\begin{figure}[t]
    \centering
    \includegraphics[width=\linewidth]{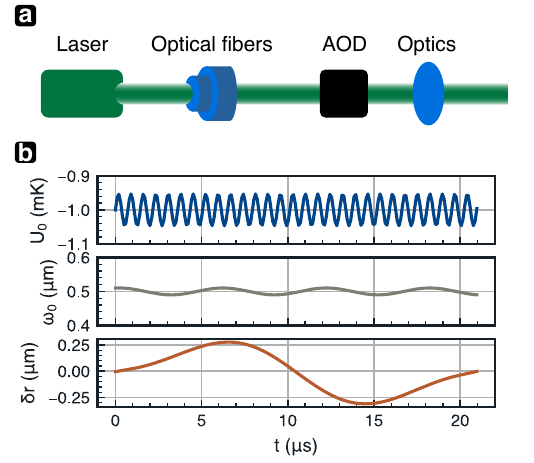}
    \caption{\textbf{Transport of neutral atoms under noisy optical tweezer.}
        (a)
        The amplitude of the laser controlling the trap depth follows a noise $\delta S$ with spectrum following a power law $S(f) = A_L / f^{1/2}$ with $A_L=10^{-11}$.
        The measured noise of the laser itself is not strong enough to affect the transport, but the beam is modified
        by the optical fibers, the AOD, and the optical elements.
        Thus, we additionally consider a sinusoidal noise $\delta U_0$ for the trap depth.
        Moreover, we consider sinusoidal noise denoted by $\delta \omega_0$ for the trap waist and $\delta r$ for the position, respectively.
        (b)
        Time-dependent pulses for trap depth $U_0$ and waist $\omega_0$ are shown in the presence of noise. The sinusoidal noise $\delta r$ for the position $r$ is depicted in the bottom plot.
    \label{fig:noise}}
\end{figure}

We then incorporate laser noise into our numerical simulation according to the description of Sec.~\ref{sec:setup}.
We summarize the setup in Fig.~\ref{fig:noise}(a) and show example pulses in panel (b).
We calculate the figure of merit for the piecewise quadratic pulse $r_{\mathrm{pq}}$ with a fixed pulse duration of $t_{\mathrm{p}}=21\,\mu$s at $T=1\, \mu$K.
Remember that this point corresponds to a minimum with $J_{\mathrm{avg}}=10^{-3}$ for $r_{\mathrm{pq}}$, as depicted in Fig.~\ref{fig:time-vs-fom-finite-T}(a).
Upon the introduction of noise, the infidelity is $J_{\mathrm{avg}}^{\mathrm{noise\,(pq)}}=0.015 \pm 0.018$.
This result is obtained by averaging over 100 simulations with different random noise instances.
The substantial standard deviation suggests that outcomes heavily rely on the specific noise instance. Consequently, scenarios may vary, ranging from instances where noise is insufficient to affect results to those where the infidelity increases by one order of magnitude.
We perform optimal control under noise and can improve the infidelity by 20\%, i.e., $J_{\mathrm{avg}}^{\mathrm{noise \,(oc)}}=0.012 \pm 0.013$.

\begin{figure}[t]
    \centering
    \includegraphics[width=\linewidth]{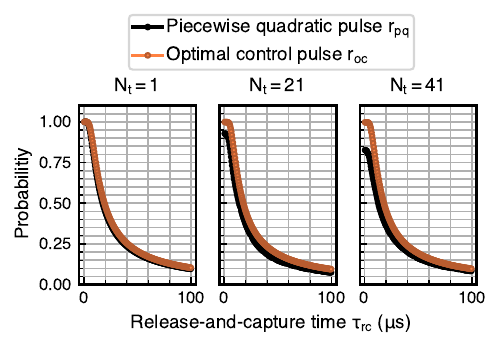}
    \caption{\textbf{Release and capture postpulse measurements.} For a temperature of $T=1\,\mu$K in the thermal state, we conduct release-and-capture measurements, i.e., the final density matrix, obtained after applying the piecewise quadratic pulse $N_t$ times, is evolved for a time $\tau_{\mathrm{rc}}$ with the trap turned off.
    The probability is computed as the accumulated probability for the wavefunction to be inside the 35th and 65th percentiles of the integrated Gaussian potential.
    We focus on the pulse with $t_{\mathrm{p}}=11\,\mu$s and compare the measurements for the piecewise quadratic pulse with the one optimized via optimal control.
    We do not observe any substantial difference between the measurements when we transport the atom only one time, i.e., for $N_t=1$.
    This absence of noticeable differences suggests that release-and-capture measurements are not sensitive to our figure of merit.
    However, when we transport the state many times, e.g., with $N_t=41$, we can observe a difference.
    }
    \label{fig:release-and-capture}
\end{figure}

In the experiment, release-and-capture measurements offer a method to estimate the temperature of the trapped atom~\cite{Hoelzl2023}. 
This provides an opportunity to integrate closed-loop
optimal control into the experiment by using the measurement from release-and-capture as a
figure of merit for the optimal control instead of $J_{\mathrm{avg}}$. To test the sensitivity
of the release-and-capture as a potential figure of merit, we simulate a release-and-capture
measurements after applying the pulse $r(t)$. Here, we examine our final density matrices obtained 
after applying the piecewise quadratic pulse $r_{\mathrm{pq}}(t)$ or the optimal pulse $r_{\mathrm{oc}}(t)$.
The sequence consists of turning off the trapping potential, allowing the wavefunction to freely
evolve, turning on the potential again after a time $\tau_{\mathrm{rc}}$, and measuring the probability
of having the atom still in the optical tweezer. While in the experiment there is a clear
path towards detecting the remaining population of atoms via fluorescence, we have to define
a corresponding figure of merit for the numerical simulations. Therefore, we calculate the
probability of the wavefunction being inside the trap as the accumulated probability to be within the 35th and 65th percentiles of the integrated Gaussian potential.
To amplify the information in the
release-and-capture measurement, we move the atom $N_t$ times back and forth in the simulation and
then measure; the backward movement is achieved by reversing the forward pulse.
For example, the atom is moved ten times back and forth between the positions of $0 \mu$m
and $3 \mu$m; with the last movement, it reaches again its final position of $3\mu$m with $N_{t} = 21$.
Looking at Fig.~\ref{fig:release-and-capture} with $N_{t} = 1$, we observe that the measurements after applying the pulses $r_{\mathrm{pq}}(t)$ and $r_{\mathrm{oc}}(t)$ are indistinguishable; the release-and-capture is not as sensitive to errors as the numerical approach.
The center panel
of Fig.~\ref{fig:release-and-capture} shows that with $N_{t} = 21$ we are now able to resolve the optimal control solution
from the piecewise quadratic pulse. Going to $N_{t} = 41$, the trend is preserved. Therefore, we propose
for a closed-loop optimal control to apply the guess pulse multiple times before running
release-and-capture measurements.

\section{Conclusion                                       \label{sec:conclusion}}

We applied open-loop optimal control to transport atoms trapped in optical tweezers over a distance of one lattice site. The wave function is modeled as a thermal state, and an analysis is conducted for various temperatures. 
While we chose the parameters for strontium-88 atoms, the entire analysis is adaptable for other species of atoms.
The optimal control is implemented using the open-source library QuOCS~\cite{Rossignolo2023}, and the dCRAB algorithm is utilized for pulse optimization~\cite{Muller2022}.
We quantified the quality of the transport using the time-averaged overlap between density matrices.
First, we simulated the transport with piecewise quadratic pulses, which then
served as a benchmark for optimal control. For the piecewise quadratic pulse,
we observed the expected behavior, i.e., local minima of the transport infidelity
matching in their periodicity the eigenfrequencies of the Gaussian potential.
Subsequently, we executed the optimal control optimization and achieved improvements of up to 89\% for the lowest chosen temperature of $1\,\mu$K.
We confirm that the results do not significantly change when we impose the constraint of having only piecewise quadratic pulse shapes in the optimal control.
When we include noise on the trap depth, waist, and position for one data point, the infidelity increases on average by two orders of magnitude. With optimal control, we can get an improvement of 10\%.
Finally, we conducted release-and-capture measurements, potentially suitable for implementation in the experiment for closed-loop optimal control optimization.
These measurements are sensitive to our figure of merit if the transport is performed several times.

The optimal control analysis can be extended to the simultaneous optimization of trap position, width, and depth.
As a first future point, the assumptions on the noise can be further refined
with experimental data toward an even more refined open-loop optimal control.
The closed-loop optimal control also has now an input in terms of speed-limit potential expectation of how to use release-and-capture measurements as a figure of merit.

\emph{Code and data availability $-$} The simulations are based on the open-source software available in~\cite{software}. 
The data to reproduce the plots and all the figures are available at~\cite{zenodo_data, figshare}.

\emph{Acknowledgments $-$}
We thank
our partners from the QRydDemo project Hans Peter B{\"u}chler,
Florian Meinert, Tilman Pfau, and J{\"u}rgen Stuhler for useful discussions.
We thank Aaron G{\"o}tzelmann, Christian H{\"o}lzl, and Christoph Tresp for providing
experimental data and discussing the experimental data and setup.
We thank Atiye Abedinnia for the discussions.
This project has received funding from the German Federal Ministry of
Education and Research (BMBF) under the funding program quantum
technologies $-$ from basic research to market $-$ with the grant QRydDemo.
We acknowledge financial support from the Italian Ministry of University
and Research (MUR) via PRIN2022 project TANQU, and the Department of
Excellence grant 2023-2027 Quantum Frontiers; from the European Union via
H2020 projects EuRyQa, TEXTAROSSA, and the Quantum Flagship project
Pasquans2, the EU-QuantERA projects QuantHEP and T-NISQ, and
from the World Class Research Infrastructure $-$ Quantum Computing
and Simulation Center (QCSC) of Padova University, and the Italian National
Centre on HPC, Big Data and Quantum Computing.
The authors acknowledge support by the state of Baden-W{\"u}rttemberg through
bwHPC and the German Research Foundation (DFG) through grant no
INST 40/575-1 FUGG (JUSTUS 2 cluster).

\bibliography{bibliography/refs}

\newpage\hbox{}\thispagestyle{empty}\newpage

\appendix

\section{Simulation approach for transport}                                        \label{appendix:code}
%

The numerical simulation for the atom transport consists of computing the dynamics of the density matrix $\rho(t)$ under the application of the time-dependent potential $V(t)$. The time evolution of $\rho$ is performed by evolving each state $\ket{\psi_i}$ from Eq.~\eqref{eq:thermal-state} independently and recombine them with their Boltzmann weights $p_i$.
The statevector evolution uses a split-operator method similar to the proposal in Ref.~\cite{Furst2014}.
The main idea of the method
is to consider the Hamiltonian of the system as a sum of the kinetic and potential parts, i.e.,
\begin{align}
    H &= K + V(x, t) \, , \qquad K = - \frac{h^2}{2m} \frac{\partial^2}{\partial x^2} \, .
\end{align}
The time-dependent Schr{\"o}dinger equation
$\mathrm{i} \hbar \frac{\partial}{\partial t} \ket{\psi} = H \ket{\psi}$ can
then be approximated via a Trotter decomposition of $K$ and $V$ for the
unitary propagator of a small time-step $\text{d}t$ as
\begin{align}
    &\exp \qty(\frac{-\mathrm{i} (K + V)  \text{d}t}{\hbar} ) \nonumber \\ 
    &\approx \qty(\frac{-\mathrm{i} V \text{d}t}{2\hbar}) \exp \qty(\frac{-\mathrm{i} K \text{d}t}{\hbar}) \exp \qty(\frac{-\mathrm{i} V \text{d}t}{2\hbar}) \, .
\end{align}
The error in the approximation depends on the commutator $[K, V]$ and with this decomposition scales with
the timestep as $\mathcal{O}(dt^3)$. The key
mechanism are Fourier transformations implemented via the FFT library to
have both components $K$ and $V$ in their diagonal eigenbasis. The exponential,
now a diagonal matrix,
becomes cheap and the FFT carries the major part of the workload~\cite{Frigo2005}. The
spatial grid is static over the evolution; no moving grid is required for the
distances considered here~\cite{Furst2014}.

\begin{figure}[h]
    \centering
    \includegraphics[width=\linewidth]{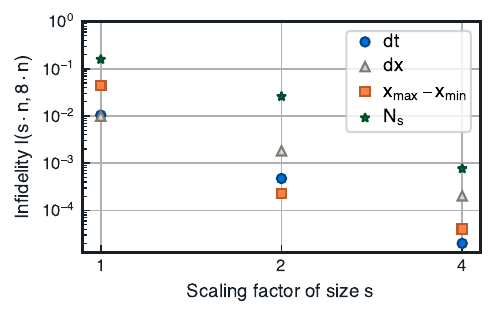}
    \caption{\textbf{Convergence plots for various parameters.} 
    For every parameter, we show the convergence by comparing simulations at size $n$, $2n$, and $4n$ where $n$ is the default value of the parameter.
    The discretization of the spatial grid $dx$ shows a scaling of $\mathcal{O}(dx^{3})$, the one of the time grid scales as $\mathcal{O}(dt^{4})$.
    For the range, we increase the limits in the space $x_{\mathrm{max}} - x_{\mathrm{min}}$ at constant discretization $dx$. The cutoff $N_s$ is shown for $T = 10\, \mu$K; we do not expect a linear scaling in log-log as the exponential distribution enters via the thermal state.
        \label{fig:conv}}
\end{figure}

\begin{figure}[h]
    \centering
    \includegraphics[width=\linewidth]{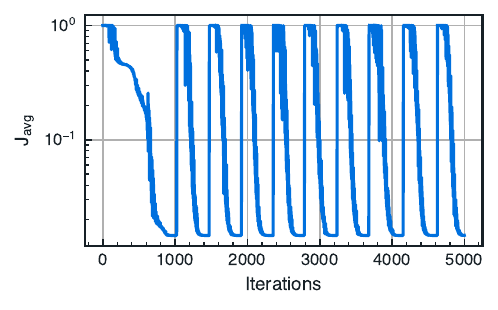}
    \caption{\textbf{Figure of merit iterations.} Optimal control iterations for $t_\text{p}=9\,\mu$s and $T=1\,\mu$K. The peaks represent the beginning of a new superiteration.
    \label{fig:example_fom_iterations}}
\end{figure}

\begin{figure*}[h]
    \centering
    \includegraphics[width=\linewidth]{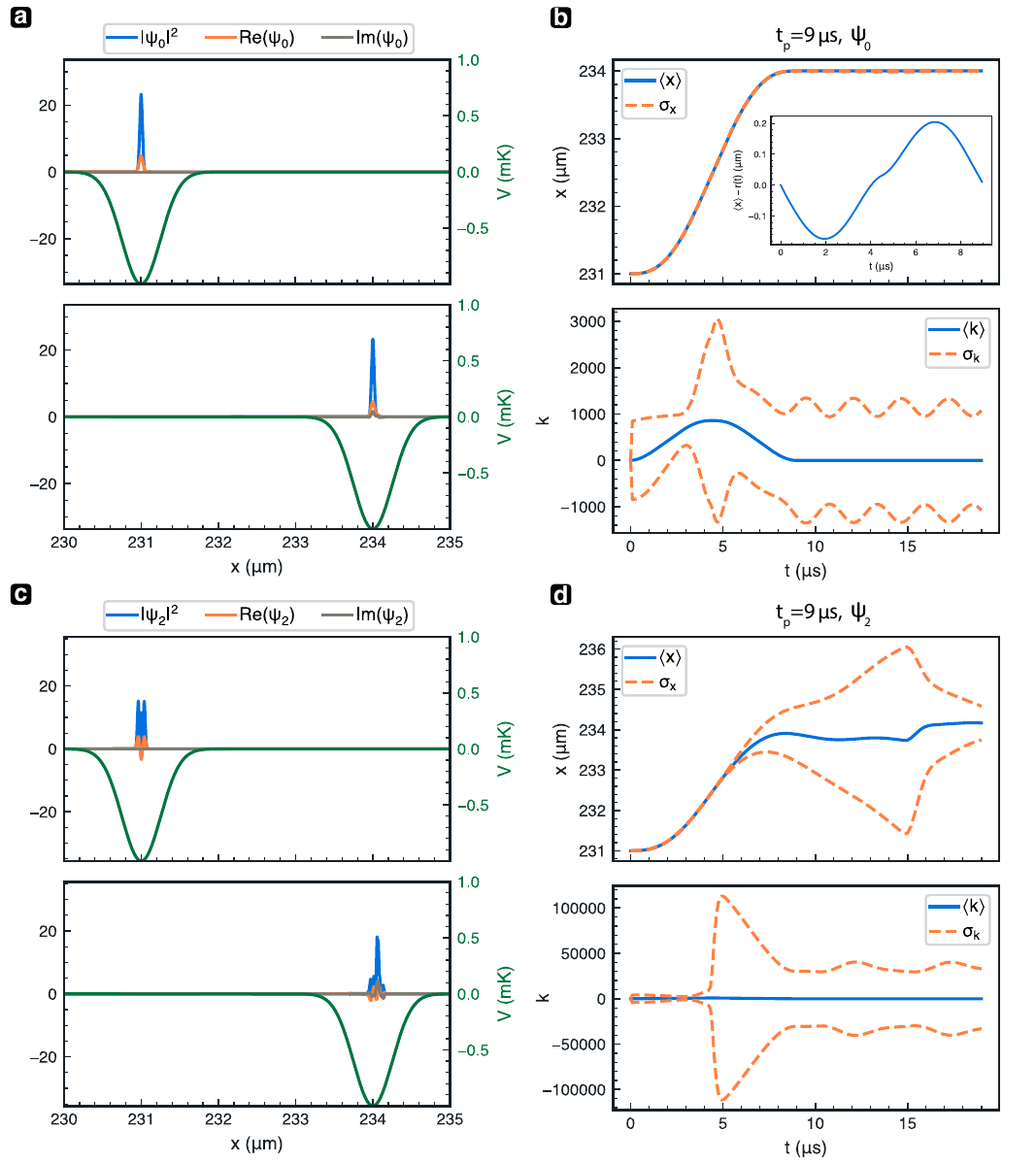}
    \caption{\textbf{Evolution for different initial states with optimal control.} (a) Ground state wavefunction before and after the evolution of duration $t_p=9\,\mu$s at $T=1\,\mu$K. (b) Expectation value of position $\langle x \rangle$ and momentum $\langle k \rangle$ for the ground state during the evolution. The inset contains the difference with the applied pulse $r(t)$ and shows the wavefunction oscillations. $\sigma_x$ and $\sigma_k$ represent the variance of position and momentum respectively. (c) Evolution of the second excited state. (d) The infidelity of the transport of $J_\text{avg} \sim 10^{-2}$ is not low enough to ensure the second excited state is not corrupted by the evolution.
     \label{fig:state_evolution}}
\end{figure*}

The convergence study targets the main parameters of the simulations, which
are the discretization in time, in space, the range in space, and the cutoff
for the finite temperature states. For each parameter, we specify the
base size of a simulation as $n$; then, we compare the convergence with
respect to simulations of a problem size of $2n$, $4n$, and $8n$, where the
latter serves as a reference point. For the convergence in discretization
in time and space, we take the piecewise quadratic scheme for a distance of $3\,\mu$m as a simulation
in case of the ground state and the infidelity is calculated based on the
overlap of the wave function
\begin{align}                                                                    \label{eq:convf1}
    \mathcal{F}(n) &= \left| \braket{\psi(8 \cdot n)}{\psi(s \cdot n)} \right|^{2},\,  s \in \{1, 2, 4 \} \, .
\end{align}
The convergence study for the extent of the grid uses the same
Eq.~\eqref{eq:convf1}, but running a release-and-capture time evolution
instead of a piecewise-quadratic transport of the atoms; the
release-and-capture is more vulnerable to the extent of the system
as there is no trapping potential during the evolution.
For the cutoff in the excited states, we use the relation
\begin{align}                                                                    \label{eq:convf2}
    \mathcal{F}(n) &= \mathcal{F}(U \rho(s \cdot n) U^{\dagger}, U \sigma(8 \cdot n) U^{\dagger}) \\ \nonumber
                   &= \left( \sum_{i=1}^{s \cdot n} \sqrt{p_{i} q_{i}} \right)^{2} \, ,
\end{align}
where $p_{i}$ and $q_{i}$ are the probabilities of the eigenstates of $\rho$
and $\sigma$, respectively. Herein, we assume that we execute the same dynamics
$U$, i.e., scenarios of optimal control finding two pulses $U$ and $V$ for different
cutoffs are beyond this analysis and beyond our interest.

\begin{table}[h]
  \begin{center}
  \caption{\textbf{Settings for convergence study.} The default settings
  in terms of the size $n$ for simulation properties to be converged.
    \label{tab:conv}}
  \begin{tabular}{@{} lc @{}}
    \toprule
    Property
    & Default size $n$ \\
    \cmidrule(r){1-1} \cmidrule(l){2-2} 
    Time discretization $dt$ & $0.5\,\mu$s \\
    Space discretization $dx$ & $0.005\,\mu$m \\
    Grid extension $x_{\mathrm{max}} - x_{\mathrm{min}}$ & $5\,\mu$m  \\
    State cutoff $N_s$ & 2      \\
    \bottomrule
  \end{tabular}
  \end{center}
\end{table}

For the convergence study, we take the setup from Tab.~\ref{tab:conv} to define
the default size $n$ of the property to be converged. The specific simulations
for each parameter lead to the convergence according to Fig.~\ref{fig:conv} and
the infidelity $\mathcal{I}$ is defined as $\mathcal{I} = 1 - \mathcal{F}$. The cutoff $N_s$ is shown for $T = 10\, \mu$K. We observe convergence with $dx$ following a scaling of $\mathcal{O}(dx^{3})$. For the time discretization the scaling is $\mathcal{O}(dt^{4})$. Neither the simulations concerning the
grid extension nor in the cutoff show a polynomial convergence. For the latter,
the behavior was expected as the exponential of the thermal state goes into the convergence.
According to the results of the convergence study, we adopt the following parameters for the simulation: a time discretization of $dt=0.1\,\mu$s, a spatial discretization of $dx=0.002\,\mu$m, a grid extension of $x_{\max} - x_{\min} = 10\,\mu$m, and a state cutoff of $N_s = 8$.
Regarding the optimization, we fix a maximum number of 5000 iterations and 30 superiterations. Moreover, the number of parameters optimized at each superiteration is fixed as $N_c = t_\text{p}/10$ if $t_\text{p} \geq 20$ and $N_c=4$ for lower times. In the end, the total number of optimized parameters depends on how many superiterations are converged. On average, the optimization converges already at the first superiteration around 1000 as shown in Fig.~\ref{fig:example_fom_iterations}.

For each statevector evolution, we store the wavefunction at each time step $t$ both in position and momentum space. We use this information to validate our results. For each initial state $\ket{\psi_i}$, we analyze plots as the ones summarized in Fig.~\ref{fig:state_evolution}.
Here, we report the example of an atom moved for $t_\text{p}=9\,\mu$s and at $T=1\,\mu$K via the optimal pulse $r_\text{oc}$. The figure of merit for this shuffling is
$J_\text{avg}\sim 10^{-2}$. Specifically, Fig.~\ref{fig:state_evolution}(a) shows the probability density for the ground state $|\psi_0|^2$ before and after the evolution. We observe that the wavefunction remains localized in the trap. Moreover, since the transport is performed with a quite high fidelity, the expectation value of the position $\langle x \rangle$ has a small variance $\sigma_x$ and the momentum reaches zero again at the end of the evolution, without considerable oscillations as shown in Fig.~\ref{fig:state_evolution}(b). However, the evolution of the second excited state $\ket{\psi_2}$ has a significantly larger standard deviation of the momentum. In Fig.~\ref{fig:state_evolution}(c-d), we see that the wavefunction $\ket{\psi_2}$ at the end is distorted and delocalized. With this argument, we can qualitatively verify the chosen figure of merit and we can understand why a figure of merit $J_\text{avg} < 10^{-2}$ is needed if we want to perform a high-fidelity transport even for the excited states.

\section{Transport distances analysis}                                        \label{appendix:transport_distances}

\begin{figure}[h]
    \centering
    \includegraphics[width=\linewidth]{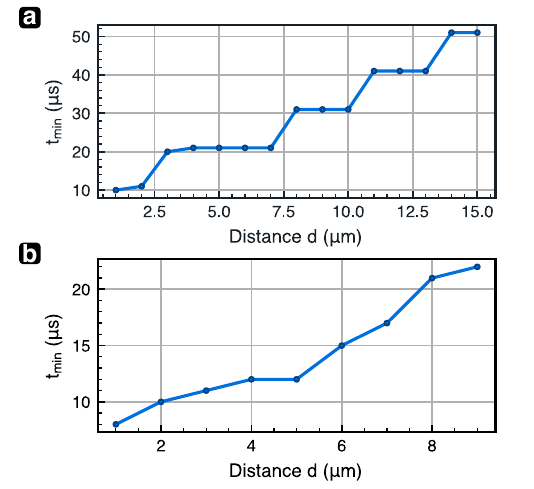}
    \caption{\textbf{Quantum speed limit for different transport distances.} (a) We report the minimum transport time with $J_{\mathrm{avg}} < 10^{-2}$ for the piecewise quadratic pulse $r_{\text{pq}}$ at $T=1\,\mu$K. (b) With optimal control, the dependence of the quantum speed limit on the distance is almost linear.
     \label{fig:distance_analysis}}
\end{figure}

In the main manuscript, we show the transport analysis for a fixed distance $d=3\,\mu$m. Here, we report the dependence of the quantum speed limit on the transport distance. In particular, we remember that the quantity $t_{\text{min}}$ is defined as the first minimum of the figure of merit which exhibits $J_{\mathrm{avg}} < 10^{-2}$.
First, we transport the atom via the piecewise quadratic pulse, and in Fig.~\ref{fig:distance_analysis}(a), we observe that the behavior shows the same periodicity of $10\,\mu$s already highlighted in Fig.~\ref{fig:time-vs-fom-finite-T}(a). For instance, with $r_\text{pq}$ we need the same time to transport the atom for $4\,\mu$m or $7\,\mu$m with a high fidelity.
Instead, for the optimal control solution, the dependence becomes almost linear as shown in Fig.~\ref{fig:distance_analysis}(b). Fluctuation of the linearity behavior may depend on the stochasticity of the optimization.

\end{document}